\documentclass{llncs}

\usepackage{makeidx}  
\usepackage{url}
\usepackage[hidelinks]{hyperref}
\usepackage{alltt}

\makeatletter
\g@addto@macro{\UrlBreaks}{\UrlOrds}
\makeatother
\usepackage{filecontents}

\begin{document}

\title{VoID-graph: Visualize Linked Datasets\\ on the Web }

\author{Luca Matteis}
\institute{
\email{lmatteis@gmail.com}
}

\sloppy
\maketitle              

\begin{abstract}
The Linked Open Data (LOD) cloud diagram is a picture that helps us grasp the contents and the links of globally available data sets. Such diagram has been a powerful dissemination method for the Linked Data movement, allowing people to glance at the size and structure of this distributed, interconnected database. However, generating such image for third-party datasets can be a quite complex task as it requires the installation and understanding of a variety of tools which are not easy to setup. In this paper we present VoID-graph (\url{http://lmatteis.github.io/void-graph/}), a standalone web-tool that, given a VoID description, can visualize a diagram similar to the LOD cloud. It is novel because the diagram is autonomously shaped from VoID descriptions directly within a Web-browser, which doesn't require any server cooperation. This makes it not only easy to use, as no installation or configuration is required, but also makes it more sustainable, as it is built using Open Web standards such as JavaScript and SVG.
\end{abstract}

\section{Introduction}
The Linking Open Data initiative started aggregating various data sets that were published using these principles, in order to make it easy for people to find the entities involved in this experiment \cite{lod}. One may say that an aggregator isn't needed in Linked Data, as you can simply follow the links. Nonetheless the idea of having relevant datasets aggregated in a single spot allowed for the creation of the famous LOD cloud diagram \cite{lodcloud}, a picture that depicted several circles of different sizes, each representing a single data set, interlinked with one another.

One important message that this diagram illustrates is that data sets that don't know of each other are still connected through third-party links. You can also understand at a glance that this information space isn't controlled or owned by any single entity, but instead is made out of various pieces contributed by different groups. Finally the diagram portrays the size of the cloud, which is an important factor for understanding the magnitude of the effort. We can therefore see that this single image can be a powerful communication method for showcasing (i) the collaborations between each entity, (ii) the distributed and unrestrained nature of such collaborations, and (iii) the number of parties involved.

Several linked datasets exist nowadays. So many in fact that it has become infeasible to display them all within a single diagram. This opens the door to the creation of smaller linking-data initiatives. These are not just single Linked Datasets, but instead efforts in creating smaller Linked Data clouds around specific data-types. An example is the Linking Open Drug Data project \cite{drugs} which aims to interconnect various open data sets about drugs, or the Linguistic Linked Open Data cloud \cite{linguistics}.

For many of these initiatives it is hard to build a diagram that exhibits the connections, the size and the context of their data. The ones that were successful in creating an LOD cloud diagram ended up writing their own scripts, tailored to their data sets. VoID-graph is a tool that makes it easy for linking-data initiatives to create their own cloud diagram and provide stakeholders a coherent visualization of the data sets and collaborations involved. In the next sections we describe the main features of the VoID-graph tool.

\section{Linked Data Visualization}

VoID-graph is an easy way to display Linked Data as a diagram. It takes advantage of the VoID vocabulary \cite{void}, a widely used way of expressing metadata about Linked Datasets. Users can simply paste their VoID description within a text area, and a basic diagram, made out of circles and links, is displayed. Everything from the parsing to the generation of the SVG diagram is executed in a Web-browser, and users can easily save the generated SVG. Compared to existing tools such as OmniGraffle\footnote{\url{https://www.omnigroup.com/omnigraffle}}, this approach is much simpler as it doesn't require the installation of anything other than a Web-browser, which is commonly available on most personal computers. VoID-graph comes as a single HTML file and can be executed by most notable Web-browsers such as Mozilla Firefox or Google Chrome without the need of browser extensions or plugins.

The various circles and arrows that make up the diagram are generated using the Web-browser's DOM manipulation capabilities. SVG nodes can be rendered in the browser's document through the use of standard JavaScript DOM APIs. JavaScript is used to parse and build the diagram; everything from the size of the text, the position of the various circles, and the length and angle of the arrows. Particularly jQuery\footnote{\url{http://jquery.com/}}, a JavaScript library, is used for styling, creating, modifying and deleting nodes as it provides an easier abstraction for DOM manipulation.

The VoID semantics are exploited to produce a diagram that is coherent with the descriptions offered. Each \texttt{void:Dataset} becomes effectively a circle and its size is derived by the \texttt{void:triples} property. Similarly \texttt{void:Linkset}, which sub-classes a dataset, is interpreted as a single arrow that connects two circles. Various properties such as \texttt{void:subset} and \texttt{void:target} are also significant, as they provide the means of linking circles together.

\section{Conclusion}

Much of the software that is used nowadays requires various complex installation and configuration steps. Specifically for web-applications it is hard to serve persistent and sustainable tools after the project's lifetime has come to an end; either the domain expires or the server becomes unresponsive. VoID-graph fixes this by utilizing what is already commonly installed and used on people's computers: a Web-browser.

VoID-graph is built using Open Web standards such as JavaScript, HTML5, CSS3 and SVG. Users can simply download the source-code and run the index.html file to utilize the tool from \url{https://github.com/lmatteis/void-graph}. Compared to other software, which requires the configuration of specific runtimes and the installation of a variety of dependencies, VoID-graph can more easily be hosted and installed on a variety of systems.

\bibliographystyle{abbrv}

\end{document}